\def \SAIT #1 #2 {{\em Mem.\ Soc.\ Astron.\ It.\/} {\bf #1}, #2}
\def \MESS #1 #2 {{\em The Messenger\/} {\bf #1}, #2}
\def \ASTRNACH #1 #2 {{\em Astron. Nach.\/} {\bf #1}, #2}
\def \AAP #1 #2 {{\em Astron. Astrophys.\/} {\bf #1}, #2}
\def \AAL #1 #2 {{\em Astron. Astrophys. Lett.\/} {\bf #1}, L#2}
\def \AAR #1 #2 {{\em Astron. Astrophys. Rev.\/} {\bf #1}, #2}
\def \AAS #1 #2 {{\em Astron. Astrophys. Suppl. Ser.\/} {\bf #1}, #2}
\def \AJ #1 #2 {{\em Astron. J.\/} {\bf #1}, #2}
\def \ANNREV #1 #2 {{\em Ann. Rev. Astron. Astrophys.\/} {\bf #1}, #2}
\def \APJ #1 #2 {{\em Astrophys. J.\/} {\bf #1}, #2}
\def \APJL #1 #2 {{\em Astrophys. J. Lett.\/} {\bf #1}, L#2}
\def \APJS #1 #2 {{\em Astrophys. J. Suppl.\/} {\bf #1}, #2}
\def \APSS #1 #2 {{\em Astrophys. Space Sci.\/} {\bf #1}, #2}
\def \ASR #1 #2 {{\em Adv. Space Res.\/} {\bf #1}, #2}
\def \BAIC #1 #2 {{\em Bull. Astron. Inst. Czechosl.\/} {\bf #1}, #2}
\def \JSQRT #1 #2 {{\em J. Quant. Spectrosc. Radiat. Transfer\/} {\bf #1}, #2}
\def \MN #1 #2 {{\em Mon. Not. R. Astr. Soc.\/} {\bf #1}, #2}
\def \MEM #1 #2 {{\em Mem. R. Astr. Soc.\/} {\bf #1}, #2}
\def \PLR #1 #2 {{\em Phys. Lett. Rev.\/} {\bf #1}, #2}
\def \PASJ #1 #2 {{\em Publ. Astron. Soc. Japan\/} {\bf #1}, #2}
\def \PASP #1 #2 {{\em Publ. Astr. Soc. Pacific\/} {\bf #1}, #2}
\def \NAT #1 #2 {{\em Nature\/} {\bf #1}, #2}

\documentstyle[twoside]{memsait}
\input epsf.sty
\begin{opening}
\title{THE SYNCHROTRON BOILER: A THERMALIZER IN SEYFERT GALAXIES}
\author{Gabriele Ghisellini$^1$, Francesco Haardt$^2$, Roland Svensson$^3$}
\institute{$^1$Osservatorio Astronomico di Brera, Via Bianchi, 46, Merate 
(Lecco), Italy\\
$^2$Dipart. di Fisica Univ. di Milano, Via Celoria, 16, Milano, Italy\\
$^3$Stockohlm Observatory, Saltsj\"obaden, Sweden}
\date{} 
\end{opening}

\begin{document}

\oddpagefooter{}{}{} 
\evenpagefooter{}{}{} 
\ 
\bigskip

\begin{abstract}
There are difficulties in understanding what keeps the plasma thermalized
in compact sources, especially during rapid variations of the emitted flux.
Particle--particle collisions are too inefficient in hot rarefied 
plasmas, and a faster process is called for.
Synchrotron absorption is such a process.
We show that relativistic electrons can thermalize in a few synchrotron 
cooling times by emitting and absorbing cyclo--synchrotron photons.
The resulting equilibrium distribution is a Maxwellian at low energies, 
with a high energy power law tail when Compton cooling is important. 
Assuming that the particles emit completely self absorbed
synchrotron radiation while they at the same time Compton scatter
ambient UV photons, we calculate the time dependent behavior of 
the distribution function, and the final high energy spectra.

\end{abstract}

\section{Set-up of the system}

\noindent
$\bullet$ In a source of dimension $R$, relativistic electrons
are injected at a rate $Q(\gamma)$ [cm$^{-3}$ s$^{-1}$] between
Lorentz factors, $\gamma_{min}$ and $\gamma_{max}$, 
corresponding to an injected
compactness, $\ell_i\equiv L_i\sigma_T/(R m_ec^3)=
4\pi R^2\sigma_T\int Q(\gamma)\gamma d\gamma /(3c)$.

\noindent
$\bullet$ A tangled magnetic field $B$ of energy density $U_B$
makes the injected particles radiate synchrotron photons.

\noindent
$\bullet$ These photons, together with photons produced externally to the 
region, interact with the electrons by the inverse Compton process.

\noindent
$\bullet$ If the particle distribution extends to a $\gamma_{max}$ of the order
of a few, the synchrotron spectrum is completely self absorbed,
and the total radiation energy density ($U_r$)
may be dominated by the externally produced photons.

\noindent
$\bullet$ The particle distribution $N(\gamma)$ is the result of 
the injection, S and IC losses, and the energy gain
due to self absorption (Ghisellini et al. 1988).

\noindent
$\bullet$ We assume that the radiation energy density, $U_r$, is 
dominated by the `hard' radiation produced by Comptonization
of an external soft photon distribution.
The latter is assumed to arise from the reprocessing
of half of the hard radiation
by cold matter in the vicinity of the active region.
We assume that the spectrum of this component is a (diluted) blackbody
with a typical temperature of 50 eV.
This is consistent with a disk--corona geometry, as discussed
by Haardt \& Maraschi (1991).


\section{Results}

\noindent
{\it Time evolution} ---
The Maxwellian shape is reached in $t\sim R/c$,
which is equal to a few synchrotron cooling times 
of the electrons with the lowest energies.

\noindent
{\it Equilibrium distributions} ---
In Fig. 1 the injection function is constant with energy,
and the different equilibrium distributions correspond to different
injected luminosities.
By increasing the injected power we increase
the relative importance of Compton cooling with 
respect to synchrotron reheating: thermal equilibrium
is then reached at lower energies (see also Ghisellini \& Svensson 1989).
In this case, the particle distribution is a Maxwellian at low
energies, with a quasi power law tail at higher energies.

\noindent
{\it High energy spectra} ---
In Fig. 2 we show the resulting Comptonized spectra corresponding
to the particle distributions shown in Fig. 1.
The spectra exhibit a quasi-exponential cut-off, similar to the
pure thermal case. 
This is true even when the injected power is large, and
the corresponding particle distribution (see Fig. 1)
is Maxwellian only at low energies.
The value of the spectral index is close to $\alpha=1$. 
{\bf The reflection bump is neglected in Fig. 2}.
Once accounting for this extra contribution, the spectra
are in agreement with the average shape shown by
Seyfert galaxies.
For small values of the injected power the spectrum is bumpy:
this is due to the small value of the optical depth and the large
temperature, resulting in well separated scattering orders.

\begin{figure}
\vskip -6.5 true cm
\epsfysize=16cm 
\hspace{2.5cm}
\epsfbox{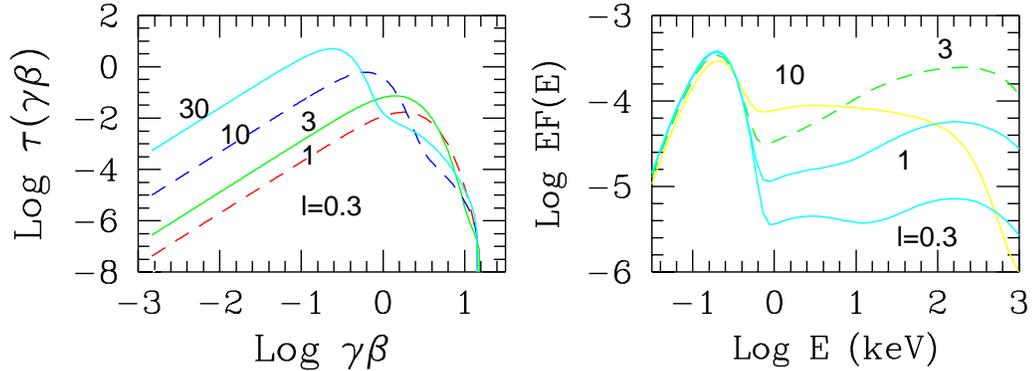} 
\vskip -5. true cm
\caption[h]{{\it ~~(Left:)}
Equilibrium particle distributions
resulting from different injected power.
Labels indicate the injected compactness.
The magnetic field is in all cases $B=1.8\times 10^4$ Gauss, $R=10^{13}$ cm,
$<\gamma>_{inj}=4.5$.  ~Fig. 2 {\it ~~(Right:)}
Comptonized spectra corresponding to four of the particle
distributions shown in Fig. 1.
Soft photons are assumed to be a blackbody at $kT=50 $ eV,
with a luminosity equal to half of the injected power.
A face-on line of sight is assumed and reflection is neglected.}
\end{figure}

\end{document}